\documentclass[12pt]{article}
\usepackage{amsmath,amssymb,amsthm,color}
\usepackage{graphicx}
\usepackage{epsfig}
\usepackage{multirow}
\usepackage{dcolumn}
\usepackage{bm}
\usepackage{float}
\usepackage{chngcntr}
\usepackage{cite}
\usepackage{amsmath} % or simply amstext
\newcommand{\Angstrom}{\textup{\AA}}
\counterwithout{figure}{section}
\renewcommand{\baselinestretch}{1.66}

\begin{document}
\title {Enrich properties of Li$^+$-based battery anode: Li$_4$Ti$_5$O$_{12}$}
\author{Thi Dieu Hien Nguyen$^{1}$\thanks{nguyenhien1901@gmail.com}, Hai Duong Pham$^{1}$, Shih-Yang Lin$^{2}$, and Ming-Fa Lin$^{1,3,4}$\thanks{mflin@mail.ncku.edu.tw}\\
%\altaffiliation[Also at ]{Physics Department, National Cheng Kung University.}%Lines break automatically or can be forced with \\
%\book mflin@mail.ncku.edu.tw (M.-F. Lin) \\ nguyenhien1901@gmail.com (T. D. H. Nguyen)
\small $^1$Department of Physics, National Cheng Kung University, Tainan 701, Taiwan\\
\small $^2$Department of Physics, National Chung Cheng University, Chiayi 62102, Taiwan\\
\small$^3$Hierarchical Green-Energy Materials Research Center, Taiwan\\
\small$^4$Quantum Topology Center, National Cheng Kung University, Tainan, Taiwan 701, Taiwan
}
%\footnotetext{$\dag~$ E-mail: mflin@mail.ncku.edu.tw}
%\footnotetext{$*$ E-mail: }
\renewcommand{\baselinestretch}{1.66}
\maketitle

\renewcommand{\baselinestretch}{1.66}

\begin{abstract}
The 3D ternary Li$_4$Ti$_5$O$_{12}$, the Li$^+$-based battery anode, presents the unusual lattice symmetry (a triclinic crystal), band structure, charge density, and density of states, under the first-principles calculations. It belongs to a large direct-gap semiconductor of ${E_g^d\sim\,2.98}$ eV. The atom-dominated valence and conduction bands, the spatial charge distribution and the atom- and orbital-decomposed van Hove singularities are available in the delicate identifications of multi-orbital hybridizations in Li-O and  Ti-O bonds. The extremely non-uniform chemical environment, which induce the very complicated hopping integrals, directly arise from the large bonding fluctuations and the highly anisotropic configurations. Also, the developed theoretical framework is very useful for fully understanding the cathodes and electrolytes of oxide compounds.
\end{abstract}

\maketitle
\section{Introduction}
The up-to-date lithium-ion batteries (LIBs; \cite{1, 111, 2, 3}) are frequently utilized in many electronic devices, such as, laptops \cite{111, 2}, cell phones \cite{111, 2}, iPods \cite{111}, and so on, being mainly due to their high capacity \cite{112, 5}, large output voltage \cite{112, 5}, long-term stability \cite{5, 6}, and friendly with the chemical environments \cite{5, 6}. The commercialized LIBs \cite{112} principally consist of a cathode (positive electrode; \cite{112}), an anode (negative electrode;\cite{112}) and an electrolyte, in which the third component  is closely related to the unusual transport of the positive lithium ions Li$^+$ between two electrodes. Very interesting, a separator membrane is designed to avoid the internal short circuit and only allow the lithium-ion to freely flow in/out the cathode and anode. The crucial mechanisms of LIBs are characterized through the unique charging and discharging process which is based on the exchange of Li$^+$-ions \cite{4}. When the charging process comes to exist, the two electrodes are connected externally to an electrical supply. The electrons are released from the cathode material and move externally to the anode that creates the charge current \cite{7}. Concurrently, the lithium ions move in the same direction internally from cathode to anode through the solid-/liquid-state electrolyte \cite{6, 7} to maintain the electric neutrality. By using this process, the external energy from the electrical supply is electrochemically stored in the battery, leading to the form of chemical energy. On the other hand, during the discharging process, electrons move in the opposite direction from anode to cathode through the external lead and thus can do the work; furthermore, lithium-ions transport back to cathode via the specific electrolyte. The discharge-process energy is very useful for the commercial purposes \cite{8}.

The experimental progress shows that two main types of negative electrodes in LIBs cover graphite and lithium titanium oxide. From the viewpoints of industry, the graphite-based anode material is very easy to be produced in LIBs, leading to the low cost. However, this systems has a serious disadvantage in the volume expansion during a lot of rapid charging and discharging processes \cite{9, 10}. While lithium titanium oxide serves as the negative electrode, it would be able to provide a long life, rapid charging, high input/output power performance, excellent low-temperature operation, a wide effective state of charge (SOC) range, and overcome the drastic volume changes in the graphitic materials \cite{10}. On the theoretical side, the graphite intercalation compounds of Li$^+$ ions/Li atoms are predicted to present the AA or AB stacking configurations \cite{11}, when the adion/adatom concentration is sufficient high or low. Furthermore, the semi-metallic/metallic behaviors, which are determined by the band overlap/the Fermi level at the conduction bands, come to exist in the Li$^+$/Li intercalation cases \cite{11, 12}. There are only few first-principles studies \cite{13} on LiTiO-related materials with the unusual superconducting properties \cite{13}. The thoroughly theoretical studies, which are conducted on their essential properties, are absent even the pure numerical calculations \cite{14}, e.g., the absence of the significant orbital hybridizations in the different chemical bonds. The critical physical/chemical/material pictures are one of the main focuses in this paper.

The previous simulation methods \cite{15}, which are based on the first-principles calculations, can provide the rich and unique phenomena, especially for the emergent 2D layered materials. For example, the systematic studies have been done for the essential properties of few-layer graphene systems \cite{16}, 1D graphene nanoribbons \cite{17}, and silicene-related materials \cite{18}, with chemical  absorptions \cite{19, 20} and substitutions \cite{21}. Such investigations clearly illustrate that the quasiparticle charges/orbitals and spins dominate all the diversified phenomena. The delicate VASP results and thorough analyses are capable of proposing the significant mechanisms/pictures in fully understanding the geometric, electronic and magnetic properties. The important multi-/single-orbital hybridizations in various chemical bondings are obtained from the optimal lattice symmetry, the atom-dominated band structures, the spatial charge densities $\&$ their variations after chemical modifications, and the atom- $\&$ orbital-decomposed density of states. Furthermore, the spin distribution configurations (the non-, ferro- $\&$ antiferro-magnetic configurations), being associated with the host and/or guest atoms, are accurately identified from the spin-split/spin-degenerate energy bands, the spin density distributions, the net magnetic moments, and the spin-projected van Hove singularities. This developed framework, which is successfully conducted on the silicene- and graphene-related systems, could be generalized to the other emergent materials, or it needs to be thoroughly test in the further investigations. It is thus expected to be very suitable for studying the rather complicated geometric and electronic properties of the main-stream Li$^+$-based batteries \cite{22}, mainly including the cathode \cite{22}, anode \cite{22} and electrolyte materials \cite{23}. In addition, the direct combinations of numerical simulations with phenomenological models would be very useful in understanding the unusual and diverse properties thoroughly, e.g., the linking of the VASP calculations \cite{16, 17, 37} and the generalized tight-binding models \cite{24} for the rich magnetic quantization \cite{25}.

This paper is mainly focused on the geometric structure and electronic properties of the Li$_4$Ti$_5$O$_{12}$-related anode material in Li$^+$-based batteries \cite{5, 6}. The first-principles calculated results cover the total ground state energy, lattice symmetry, various Li-O and Ti-O, the atom-dominated band structure, the spatial charge density distribution, and the atom- $\&$ orbital-projected van Hove singularities. The spin-dependent behaviors, the spin-split electronic states, the net magnetic moment and the spin density distributions, will be fully test whether they could survive in the ternary transition-metal-atom compound. Such physical properties will play important roles to achieve the critical multi-orbital hybridizations in three kinds of chemical bonds. The analysis difficulties lie in the very complicated orbital-decomposed density of states, being supported by the electronic energy spectrum and carrier density. The theoretical predictions, which are conducted on the optimal geometry, the occupied valence state and the energy gap $\&$ whole energy spectrum, could be verified from the high-resolution measurements of X-ray diffraction/low-energy electron diffraction (LEED; \cite{26}), angle-resolved photoemission spectroscopy (ARPES; \cite{27}), and scanning tunneling microscopy (STM; \cite{28}).

\section{Theoretical simulation methods}

The rich and unique geometric structures and electronic properties of the three-dimensional $Li_4Ti_5O_{12}$ compound are thoroughly investigated through the density functional theory (DFT; \cite{29}) implemented by Vienna ab initio simulation package (VASP; \cite{30}). The many-particle exchange and correlation energies, which mainly arise from the electron-electron Coulomb interactions, are calculated from the Perdew-Burke-Ernzerhof (PBE; \cite{31}) functional under the generalized gradient approximation. Furthermore, the projector-augmented wave (PAW; \cite{32, 33}) pseudopotentials are able to characterize the electron-ion interactions. It is well known that these two kinds of intrinsic interactions have no exact formulas \cite{34}, i.e., it is very difficult to express the single- and many-particle Hamiltonian in the analytic form \cite{34}. In general, the plane waves, with the kinetic energy cutoff of 520 eV, is chosen as a complete set \cite{34}; therefore, they are very reliable and suitable for evaluating Bloch wave functions and electronic energy spectra. The first Brillouin zone is sampled by 3 $\times$ 3 $\times$ 3 and 12 $\times$ 12 $\times$ 12 k-point meshes within the Monkhorst-Pack scheme for geometric optimizations and electronic structures, respectively. Such points are enough in obtaining the reliable orbital-projected density of states, spatial charge distributions, and spin density configurations. The convergence for the ground-state energy is 10$^{-5}$ eV between two consecutive simulation steps, and the maximum Hellmann-Feynman force acting on each atom is less than 0.01 eV/$\Angstrom$ during the ionic relaxations.

By the delicate VASP calculations and detailed analyses are conducted  on certain physical quantities, such as \cite{30}, the atom-dominated band structures, the spatial charge densities, the atom- $\&$ orbital-decomposed van Hove singularities, the spin  distribution configurations, the spin-split or spin-degenerate energy bands, and the net magnetic moments. As a result, the critical pictures, the multi- and/or single-orbital hybridizations in chemical bonds and the spin configurations due to different atoms, could be achieved under the concise scheme. Such viewpoints will be very useful in fully comprehending the diversified physical/chemical/material phenomena. The theoretical framework has been successfully in the systematic investigation on the geometric, electronic, and magnetic properties of few-layer 2D graphene systems \cite{35}, 1D graphene nanoribbons \cite{36} and 2D silicene-relayed material. Very interesting, the chemical modifications through the adatom chemisorptions and guest-atom substitutions can greatly diversify the various fundamental properties \cite{37}. Apparently, the developed viewpoints are available in other condensed-matter systems. For example, they should be suitable in thoroughly exploring the diverse phenomena of $Li^{+}$-based battery anode/cathode/electrolyte materials \cite{1}/\cite{111}/\cite{2}.

The numerical simulations might be available in combining with the phenomenological models. For example, the first-principles calculations on band structures could be well fitted by the tight-binding model/the effective-mass approximation if the electronic energy spectra across the Fermi level are not so complicated \cite{37}. This viewpoint has been very successful in fully exploring the diversified essential properties in few-layer graphene systems, e.g., the diverse magnetic quantization of AA- \cite{38, 39, 40}, AB- \cite{40, 41}, ABC- \cite{42} and AAB-stacked graphenes under the generalized  tight-binding model with any external fields. Such model is closely related to the parameterized Hamiltonian; that is, the various hopping integrals, which are due to the different orbital hybridizations, need to be taken into account for its diagonalization. Apparently, they play critical roles in expressing the suitable and reliable Hamiltonian. Specifically, the current material of Li$_4$Ti$_5$O$_{12}$ presents an extremely non-uniform chemical environment in a primitive unit cell (discussed later in Fig. 4.1); therefore, it might be very difficult to achieve a reliable tight-binding model with the various chemical bondings.

\section{Rich geometric symmetries of 3D Li$_4$Ti$_5$O$_{12}$ compound}

The three-dimensional Li$_4$Ti$_5$O$_{12}$ compound exhibits the unusual geometric symmetries. One of meta-stable configurations, which possesses the smallest unit cell, is chosen for a model study. This corresponds to a triclinic structure, as clearly illustrated in Fig. 4.1. A primitive unit cell has 42 atoms (8-Li, 10-Ti and 24-O atoms), the lattice constants of $a$=5.288 $\Angstrom$, $b$=9.532 $\Angstrom$ and $c$=9.932 $\Angstrom$, and the titled angles of $\alpha$=73.122$^\circ$, $\beta$=78.042$^\circ$, and $\gamma$=78.913$^\circ$ about $\hat x$, $\hat y$ and $\hat z$, respectively. The space-group symmetries belong to Hermann Mauguin (P1), Hall ($-$P1), and point group ($\bar{1}$). Apparently, there exist the highly anisotropic and the extremely non-uniform chemical environment. The projections of this 3D material on the different planes present the diverse atom arrangements, such as, those of (a) (100), (b) (010), (c) (001)(Figs. 4.2(a)-4.2(c)), where (Li, Ti, O) atoms are, respectively, represented by the blue, red, and green balls. The real-space lattice is available in getting the reciprocal lattice, where the band structure is done along the paths of high-symmetry points: X-$\Gamma$-Y$|$ L-$\Gamma$-Z$|$N-$\Gamma$-M$|$R-$\Gamma$ (their definitions in (Fig. 4.3)).

The unique chemical bonds, which survive in a primitive unit cell will determine the fundamental properties. They cover Li-O and Ti-O  bonds, in which their total numbers are 48 and 52, respectively. According to the delicate first-principles calculations in Table. 4.1,
their bond lengths might lie in a wide range of  $\sim$1.988-2.479 $\Angstrom$ and 1.757 -2.210$\Angstrom$. This diversified phenomenon will be directly reflected in the distribution width of the spatial charge density (Fig. 4.5). Both Li-O and Ti-O bonds exhibit the sufficiently large modulations, clearly indicating the non-uniform bonding strengths. This might lead to the easy intercalation of the Li$^+$ ions and thus the structural transformation to another meta-stable geometric configuration. Furthermore, all the chemical bonds are generated by the multi-orbital hybridizations (discussed later in orbital-dependent density of states by Fig. 4.6).  A lot of various orbital hybridizations come to exist in this anode material, and so do the hopping integrals in the tight-binding model \cite{43}. The theoretical predictions on the optimal geometric structures could be verified from the experimental examinations. Both X-ray diffractions \cite{44} and low-energy electron diffractions (LEED; \cite{26, 46}) are suitable in measuring the 3D lattice symmetries, but not scanning tunneling microscopy (STM, \cite{47}) and tunneling electron microscopy (TEM; \cite{28, 47}) for the nano-scaled top- and side-view structures, respectively. For the 3D Li$_4$Ti$_5$O$_{12}$ material, they could be utilized to thoroughly explore the diversified lengths in Li-O and Ti-O bonds. This will be very useful in indirectly identifying the complicated multi-orbital hybridizations in all the chemical bonds.

\section{Rich and unique electronic properties}

The 3D ternary compound of Li$_4$Ti$_5$O$_{12}$ exhibits the unusual electronic states. The first-principles band structure, as indicated in Fig. 4.4(a), is very sensitive to the changes of wave vector. Consequently, it is highly anisotropic. Apparently, the occupied electron energy spectrum is asymmetric to the unoccupied hole one about the Fermi level (${E_F-0}$), mainly owing to the  different multi-orbital hybridizations in the non-uniform chemical bonds (Fig. 4.1). The highest occupied valence-band state and the lowest unoccupied conduction-band one determines energy gap of ${E_g^d\sim\,2.98}$ eV (a red perpendicular arrow); furthermore, they come to exist at the same wave vector, leading to a direct-gap semiconductor. This band gap is just equal to the optical threshold absorption frequency; that is, the optical spectroscopies \cite{48} are very suitable in examining the theoretical prediction on ${E_g^d}$. $E_g$ is too wide to generate the spin-split electronic states across the Fermi level, where this phenomenon is supported by the spin-degenerate valence and conduction bands, respectively, below and above $E_F$ with a sufficient great energy spacing. As a result, the net magnetic moment is zero and the  spin-dependent interactions could be ignored in this ternary compound. There exist a plenty of valence and conduction energy subbands because of many atoms and outer orbitals in a large unit cell. Generally speaking, the energy dispersions are weak, but significant. Their widths might lie in the range of ${\sim\,0.08-0.31}$ eV. The different energy subbands would present the non-crossing, crossing and anti-crossing behaviors. It should be noticed that the last phenomenon will appear when two neighboring subbands possess the comparable independent components within the anti-crossing region \cite{49}, e.g., the almost competitive amplitude of the same mode in two anti-crossing Landau levels \cite{49}.

The atom dominances, which are represented by the ball radius, are very useful in understanding the critical roles of chemical bonding on the rich electronic states, such as, the contributions to electronic states for the lithium, titanium and oxygen atoms through the blue, red and green balls, respectively. It is very difficult to observe the obvious contributions due to the Li atoms, in which only small blue circles are revealed in the whole range of valence and conduction subbands (Fig. 4.4(b)). This unique result clearly illustrates that  the chemical bonding strengths (the hopping integrals of neighboring atoms; \cite{24, 56}) of the 48 Li-O bonds (Figs. 4.2 and 4.3) should present a very wide range because of the large-fluctuation of their lengths; furthermore, each Li atom only makes the single-orbital contribution of 2s. On the other side, the Ti-contributions are observable and very important within the valence- and conduction-band ranges of ${-5.50}$ eV${\le\,E^v\le -1.49}$ eV and ${1.49}$ eV${\le\,E^c\le 4.0}$ eV, respectively, especially for the latter. Apparently, the unoccupied electronic states are dominated by the titanium atoms, but not the oxygen ones. The opposite is true for the oxygen contributions. Such phenomenon might be associated with the simultaneous existence of many O-related chemical bonds (100), with the deeper (2s, 2p$_x$, 2p$_y$, 2p$_z$)-orbital energies. In short, the atom-dominated electronic structure could provide the partial information about the critical multi-orbital hybridizations.

The theoretical predictions, which are conducted on the wave-vector-dependent band structures below $E_F$, could be verified from the high-resolution ARPES \cite{27}. The previous many measurements have clearly identified the low-energy valence bands of layered graphene systems, being only initiated from the K/K$^\prime$ valleys \cite{50}, such as, the linear Dirac-cone structure in twisted bilayer graphenes (the Moire superlattices with very large unit cells; \cite{51}) $\&$monolayer graphene \cite{52}, the parabolic/parabolic and linear dispersions in bilayer/trilayer AB stackings \cite{52}/\cite{53}, the linear, partially flat and Sombrero-shaped bands in tri-layer ABC stacking \cite{53, 55}, and the semi-metallic property in bulk Bernal graphite \cite{53}. These diversified electronic energy spectra are identified to only arise from the pure and unique interlayer ${2p_z}$-${2p_z}$ orbital hybridizations in the normal/enlarged honeycomb lattices, according to a good consistence between the first-principles method \cite{56} and the tight-binding model \cite{24, 56}. The similar experimental measurements are available in examining the main features of the 3D ternary Li$_4$Ti$_5$O$_{12}$ compound, such as, the sensitive dependence on wave vector, the large band gap, the highly asymmetric electron and hole energy spectra, the weak, but significant energy dispersions, and the frequent non-crossing/crossing/anti-crossing behaviors. Their verifications are very helpful in solving the critical orbital hybridizations of chemical bonds.

The spatial charge distributions on the different chemical bonds, as clearly indicated in Figs. 4.5(a)-4.5(h), are capable of providing the first-step orbital hybridizations, being further examined by the delicate atom- and orbital-decomposed van Hove singularities (discussed later in Fig. 4.6). The optimal geometry, with a triclinic unit cell (Fig. 4.2), show the LiO$_6$, TiO$_4$ and TiO$_6$ chemical bonds, illustrating the highly non-uniform chemical environment. Their distinct bond lengths (Table 4.1) only directly the diversified chemical bonding strengths (the diverse charge density distributions). That is to say, the spatial carrier densities, being closely related to electron orbitals of each atom, are very sensitive to the changes of the nearest-neighbor distances. As for the shortest Li-O bonds (1.988 $\Angstrom$ in Fig. 4.5(d)), the only 2s-orbital only provides a dilute charge density around the lithium atom (Fig 4.5(d)), in which the effective distribution range is about 0.52 $\Angstrom$ as measured from the deep blue region of the Li$^+$-ion core to the  light one of the extended 2s-orbital. Apparently, the two 1s orbitals are independent of the critical orbital hybridizations in the Li-O bonds. There exists an obvious overlap of distinct orbital charges in the range of 1.20-1.60 $\Angstrom$ the distance with the O-core. Furthermore, the significant O-orbitals, being indicated by the light green and yellow colors, are deduced to arise from the (2p$_x$, 2p$_y$, 2p$_z$) ones. However, the O-dependent 2s orbitals are far away from that of the lithium atom; therefore, their contributions to the chemical bonds are almost negligible. With the decreasing of Li-O length (2.479 $\Angstrom$ and 2.249 $\Angstrom$ in Figs. 4.6(b) and 4.6(c), respectively), the charge overlap behavior between Li and O atoms is obviously increased. Very interesting, the multi-orbital hybridizations of 2s-(2p$_x$, 2p$_y$, 2p$_z$) in Li-O bonds present the diverse hopping integrals \cite{56}.

There are more Ti-O chemical bonds and quite strong bonding strengths (Figs. 4.5(e)-4.5(g)), compared with those of Li-O bonds. The effective distribution range of Ti atom is somewhat higher than that of O one because of the large atomic number. Most important, it could be classified into the heavy red, light red, and yellow-green regions, which, respectively, corresponds to (3s, 3p$_x$, 3p$_y$, 3p$_z$), (4s), and (3d${_{x^2-y^2}}$, 3d${_{xy}}$, 3d${_{yz}}$, 3d${_{zx}}$, 3d${_{z^2}}$). Apparently, the first ones do not take part the important multi-orbital hybridizations in the Ti-O bonds. The deformed carrier distributions between Ti and O atoms, as shown in the distinct bond-length cases, clearly illustrate the diversified charge transfers from the former to the latter. The extremely non-uniform chemical bondings even for the similar bonds will induce the high barriers in generating the suitable hopping integrals for the phenomenological models, e.g., the reliable parameters in the tight-binding model \cite{56, 57}.

The atom- and orbital-projected density states, as clearly indicated in Figs. 4.6(a)-4.6(d), are able to provide the full information on the significant multi-orbital hybridizations on the Li-O and Ti-O chemical bonds. Apparently, the density of states per unit cell vanishes with a large band gap of ${E_g^d\sim\,2.98}$ eV  (Figs. 4.4(a)-4.4(d)), and its valence and conduction spectra are highly asymmetric to each other. Whether a very wide gap will induce the difficulties of experimental observations could be test in the further STS measurements (discusses later; \cite{59}). There exist a lot of shoulders and asymmetric/symmetric peaks. Such van Hove singularities principally originate from the band-edge states of energy subbands, with the local minimum, maximum, saddle and almost dispersionless points in the energy wave-vector spaces. They might appear in between the high-symmetry points. It is well known that the 3D parabolic energy dispersions can create the square-root dependences \cite{60}. Generally speaking, the titanium and oxygen atoms, respectively, dominate the density of states in the energy range of ${-5.50}$ eV${\le\,E^v\le -1.49}$ eV and ${1.49}$ eV${\le\,E^c\le 4.0}$ eV, respectively (the violet and orange curves). Furthermore, the lithium atoms only make minor contributions in the whole valence and conduction energy spectrum (the green curve in Fig. 4.6(b)) through the only 2s orbital, but not two core-level 1s orbitals.

The important contributions of various orbitals, which mainly come from (I) Li-2s orbital (the blue curve in Fig. 4.6(b)), (II) Ti-(4s, $3d{_{x^2-y^2}}$, $3d{_{xy}}$, $3d{_{yz}}$, $3d{_{zx}}$, $3d{_{z^2}}$) orbitals (the red, heavy red, purple, brown, light green; light blue curves in Fig. 4.6(c)), and (III) O-(2s, $2p_x$, $2p_y$, $2p_z$) orbitals (the purple, pink, green and orange curves in Fig. 4.6(d)) are worthy of a closer examination. The enlarged spectral scale, which is done for lilthium atoms, is able to provide the clear van Hove singularities, i.e., their number, energies, intensities and forms are clearly illustrated in Fig. 4.6(b). These main features are similar to those of ($2p_x$, $2p_y$, $2p_z$)-decomposed density of states for oxygen atoms (Fig, 4,6(d)). The similarities, which cover the whole energy spectrum (e.g., ${-6.0}$ eV${\le\,E\le\,4.0}$ eV), only reflect the very important 48 Li-O chemical bonds with the multi-orbital hybridizations (Figs. 4.1-4.2). It should be noticed that the O-2s orbitals make the almost zero contribution in the specific energy range. The principal reason might be that they belong to the fully occupied states under the spin-up and spin-down configurations; therefore, any dangling bonds/chemical bondings are forbidden. The effective width of energy spectrum, being closely relate to the Li-O Chemical bonds, are more than 10.0 eV. This unusual characteristic is attributed to the complicated 2s-($2p_x$, $2p_y$, $2p_z$) chemical bondings (the multi-orbital hopping integrals and on-site Coulomb potentials; \cite{56}, and the large bond-length modulations (1.988-2.479 $\Angstrom$; the easily modulated hopping integrals).

In addition to the Li-O bonds, the oxygen atoms also make the  significant contributions to the Ti-O bonds, especially for the valence-state energy spectrum. The obvious evidences are revealed in the similar van Hove singularities in terms of the orbital-decomposed special forms, energies, and numbers. Concerning the transition-metal titanium atoms  there exist unique six orbitals (4s, 3d${_{x^2-y^2}}$, 3d${_{xy}}$, 3d${_{yz}}$, 3d${_{zx}}$, 3d${_{z^2}}$) orbitals which take part in the chemical bondings of  oxide compounds, as clearly illustrated in Fig. 4.6(c). The orbital-dependent contributions are almost comparable in the whole energy spectrum except for the small 4s-orbital density of state with the conduction-band range (the pink curve). Very interesting, the initial conduction-/valence-state energy spectrum is mainly determined by the titanium and oxygen atoms. In short, the  Li-O and Ti-O  chemical bonds are deduced to have the multi-orbital hybridizations of 2s-(2p$_x$, 2p$_y$, 2p$_z$), (4s, 3d${_{x^2-y^2}}$, 3d${_{xy}}$, 3d${_{yz}}$ and 3d${_{zx}}$, 3d${_{z^2}}$)-(2p$_x$, 2p$_y$, 2p$_z$), respectively.

The high-resolution STS measurement is the most efficient technique in examining the various van Hove singularities due to the band-edge states. Up to date, a rather weak tunneling quantum current could be measured in the very accurate way; furthermore, its differential conductance of dI/dV is deduced to be approximately proportional to the density of states. For example, such experiments have been successful in identifying the significant coupling effects in few-layer graphene systems, e.g., an almost symmetric V-shape structure vanishing at  the Fermi level for monolayer (a zero-gap semiconductor; \cite{61}), the logarithmically symmetric peaks in twisted bilayer graphenes \cite{62}, a gate-voltage-induced energy gap for bilayer AB and tri-layer ABC stackings \cite{53, 62}, a delta-function-like peak centered about $E_F$ in
tri-layer and penta-layer ABC stackings \cite{64}, a sharp dip structure near $E_F$ combined with a pair of square-root peaks under tri-layer AAB stacking (a narrow-gap semiconductor with constant-energy loops; \cite{65}). The further experimental examinations, which are very suitable for the 3D ternary Li$_4$Ti$_5$O$_{12}$ compound, are require to detect a large band gap of ${E_g^d\sim\,2.98}$ eV, a lot of asymmetric/symmetric peaks $\&$ broadening shoulders, the high asymmetry of electron $\&$ hole energy spectrum, and their different widths. The measured results, being combined with the  theoretical predictions on van Hove singularities, might be very useful in comprehending the complicated multi-orbital hybridizations of Li-O and Ti-O.

The first-principles results could be utilized to establish the phenomenological model, the tight-binding model, as discussed in Sec. %??
But for the 3D Li$_4$Ti$_5$O$_{12}$ compound \cite{56}, there exist a lot of different chemical bonds in a primitive unit cell (100 Li-O and Ti-O bonds Figs. 4.1-4.2), very complicated valence and conduction subbands with a large band gap of ${E_g=2.98}$ eV (Figs. 4.4(a)-4.4(d)), non-homogeneous spatial charge densities in diverse chemical bonds (Figs. 4.5(a)-4.5(g)), and three kinds of multi-orbital hybridizations in van Hove singularities (Figs. 4.6(a)-4.6(d)). Such critical factors are almost impossible to be included in the parameterized tight-binding model.  That is to say, the extremely non-uniform hopping integrals, which are associated with the various orbital hybridizations, are rather difficult to be achieved to fit the main features of the first-principles band structure. Apparently, how to obtain a reliable tight-binding model in Li$^+$-based anode \cite{66}, cathode \cite{66} and electrolyte materials \cite{66} will become an open issue, since it is relatively easy in understanding the essential physical/material/chemical properties \cite{67} from the concise pictures

\section{Concluding remarks}

The theoretical framework, which is based on the first-principles calculations \cite{37}, is developed for the essential properties of the 3D ternary Li$_4$Ti$_5$O$_{12}$ compound. The critical multi-orbital hybridizations in Li-O and Ti-O chemical bonds are delicately identified from the atom-dominated valence and conduction bands, the spatial charge density, and the atom- and orbital-decomposed density of states. Their highly anisotropic and non-uniform characteristics in a large unit cell clearly indicate show that the reliable tight-binding model, with the various hopping integrals and on-site Coulomb potentials, is almost impossible to achieve for the simulation of the VASP band structure. The similar developments could be generalized to the Li$^+$-based battery cathode \cite{68}, anode \cite{69} and electrolyte materials \cite{70}. Especially, the rich oxide compounds are worthy of the systematic investigations for the diversified physical/chemical/material science phenomena.

The current anode material, with the smallest unit cell of 42 atoms, is a triclinic structure in the titled three axes. There exist 100 chemical bonds, in which the very strong covalent bondings make this system present a large direct gap of ${E_g^d\sim\,2.98}$ eV, being equal to the optical threshold absorption frequency. The high-resolution measurements of optical spectroscopies \cite{71} are expected to be very efficient in examining the semiconducting behavior. A lot of valence and conduction bands are highly asymmetric to each other about the Fermi level, in which their energy dispersions are weak along any directions. Generally speaking, the whole band structure, being related to the critical chemical bondings, lie in the range of ${-6.0}$ eV${\le\,E^v\le\,4.0}$ eV. It also reveals the frequent subband non-crossings, crossings and anti-crossings \cite{49}. Moreover, the band-edge states, the critical points in the energy-wave-vector spaces, appear as the van Hove singularities in density of states. The atom- and orbital-dependent special structures, the asymmetric/symmetric peaks and broadening shoulders, are successful in identifying the important multi-orbital hybridizations, such as, the 2s-(2p$_x$, 2p$_y$, 2p$_z$),  and (4s, 3d${_{x^2-y^2}}$, 3d${_{xy}}$, 3d${_{yz}}$, 3d${_{zx}}$, 3d${_{z^2}}$)-(2p$_x$, 2p$_y$, 2p$_z$) in  Li-O and Ti-O bonds, respectively. The critical mechanisms are also supported by the spatial charge density distribution The diverse covalent bondings are partially supported by the atom-created energy bands and charge density distributions. Very interesting, in addition to Li$_4$Ti$_5$O$_{12}$ compound, there exist other 3D LiTiO-related ternary materials, e.g., LiTi$_2$O$_{4}$ \cite{72}, Li$_2$Ti$_2$O$_{4}$ \cite{73}, and Li$_7$Ti$_5$O$_{12}$ \cite{73}. Such unusual condensed-matter systems are examined to have the diverse geometric symmetries \cite{73, 74} and thus exhibit the quite different electronic \cite{73}, optical \cite{73}, and superconducting properties. This current study clearly shows the very large variations of chemical bond lengths. The relatively easy modulations on chemical bonds might be helpful in solving the open issue: the structural transformation between two meta-stable configurations during the charging or discharging processes of Li$^+$-based batteries.\\
  Table 4.1: The various chemical bond lengths in Li$_4$Ti$_5$O$_{12}$, in which the total number of Li-O and Ti-O bonds is 100.
            \begin{center}
             \begin{tabular}{c c c c} \hline
              \textbf{No. of bonds}&\textbf{Atom-Atom}&\textbf{Bond length ($\Angstrom$)} \\ \hline
                 48&Li-O&1.9877-2.4791\\
                 52&Ti-O&1.7571-2.2098\\

            \end{tabular}
            \end{center}

  \newpage
Figure Captions
 \setcounter{figure}{0}
\begin{figure}[!ht]
         \centering
          \includegraphics[scale=.8]{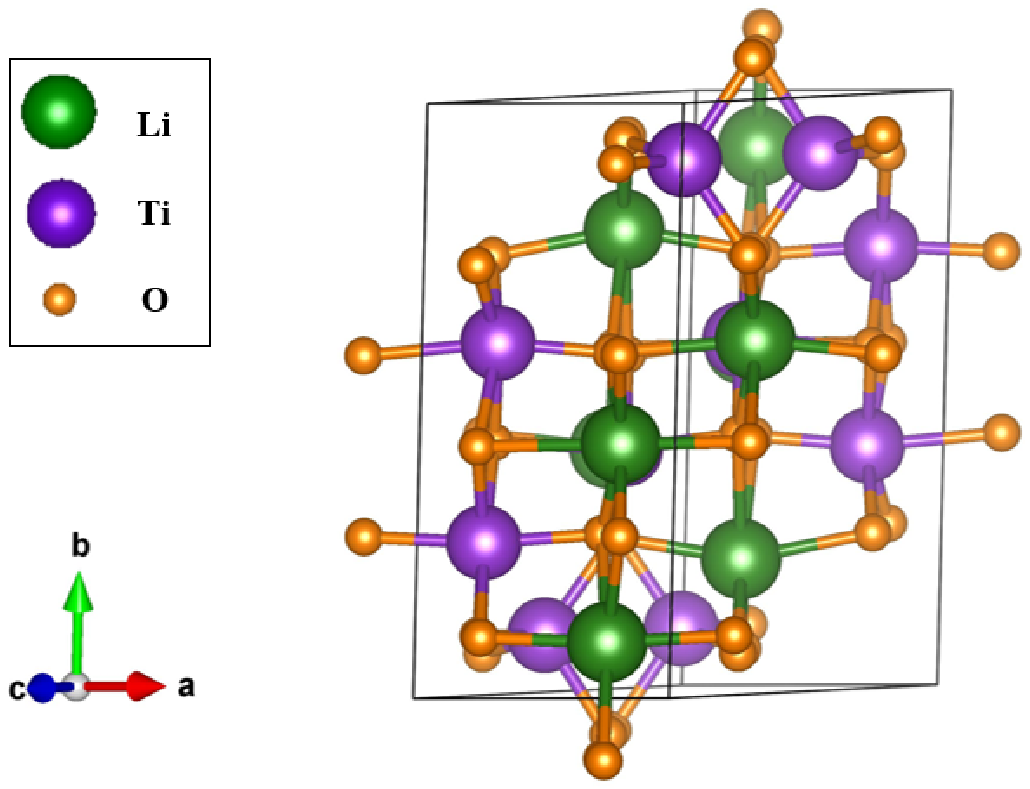}

\end{figure}

Figure 4.1: The optimal geometry of Li$_4$Ti$_5$O$_{12}$, with triclinic symmetry, where a primitive unit cell has 42 atoms, the lattice constants of a=5.288 $\Angstrom$, b=9.532 $\Angstrom$ and c=9.932 $\Angstrom$; the titled angles of $\alpha$=73.122$^\circ$, $\beta$=78.042$^\circ$, and $\gamma$=78.913$^\circ$.
\newpage
\vfill
\begin{figure}[!ht]
            \centering
            \includegraphics[scale=.75]{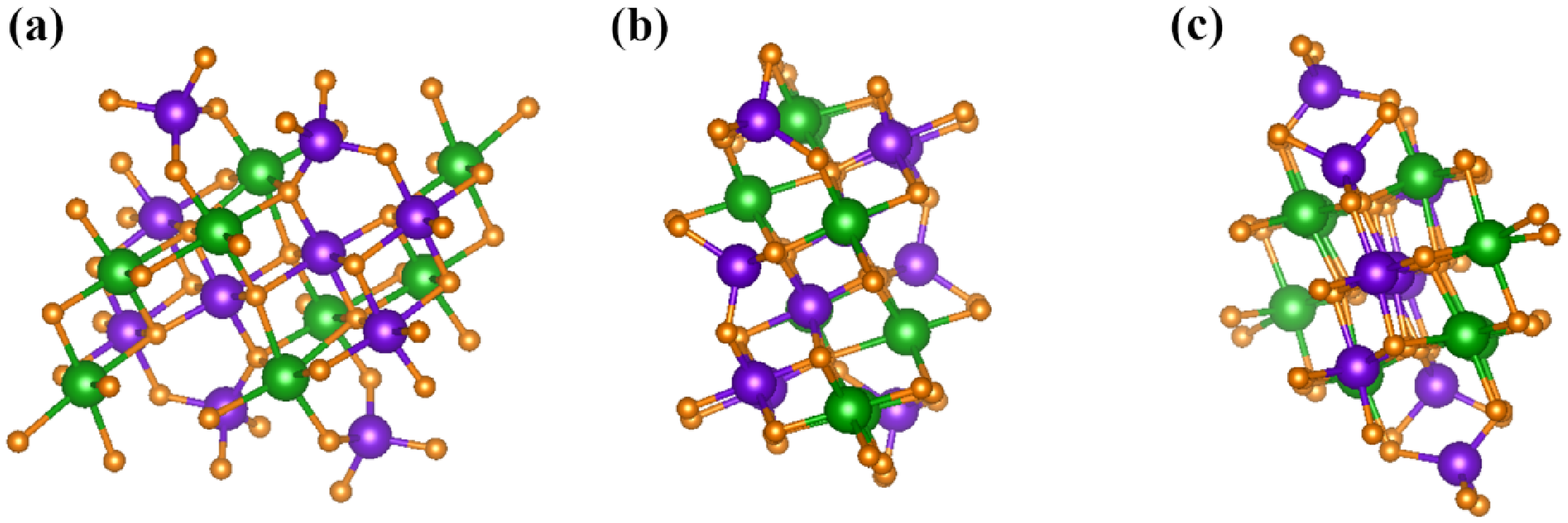}
\end{figure}

Figure 4.2: The geometric structure of Li$_4$Ti$_5$O$_{12}$ along the different projections: (a) (100), (b) (010), (c) (001), in which Li, Ti and O atoms are, respectively, represented by the green, violet, and orange balls.
\begin{figure}[!ht]
            \centering
            \includegraphics[scale=.6]{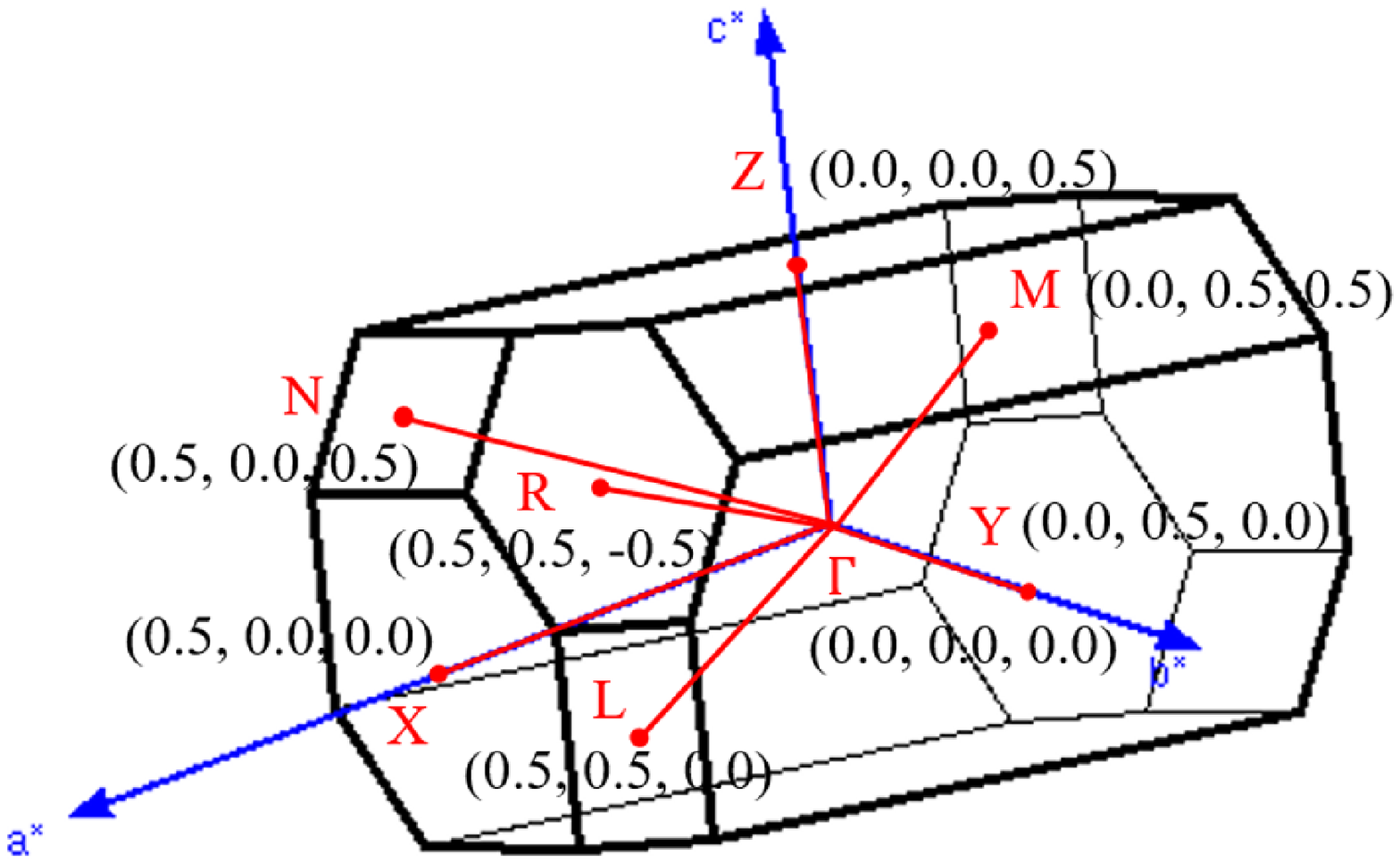}

\end{figure}

Figure 4.3: The first Brillouin zone with the paths of high-symmetry points.
\vfill
\begin{figure}[!ht]
            \centering
            \includegraphics[scale=0.85]{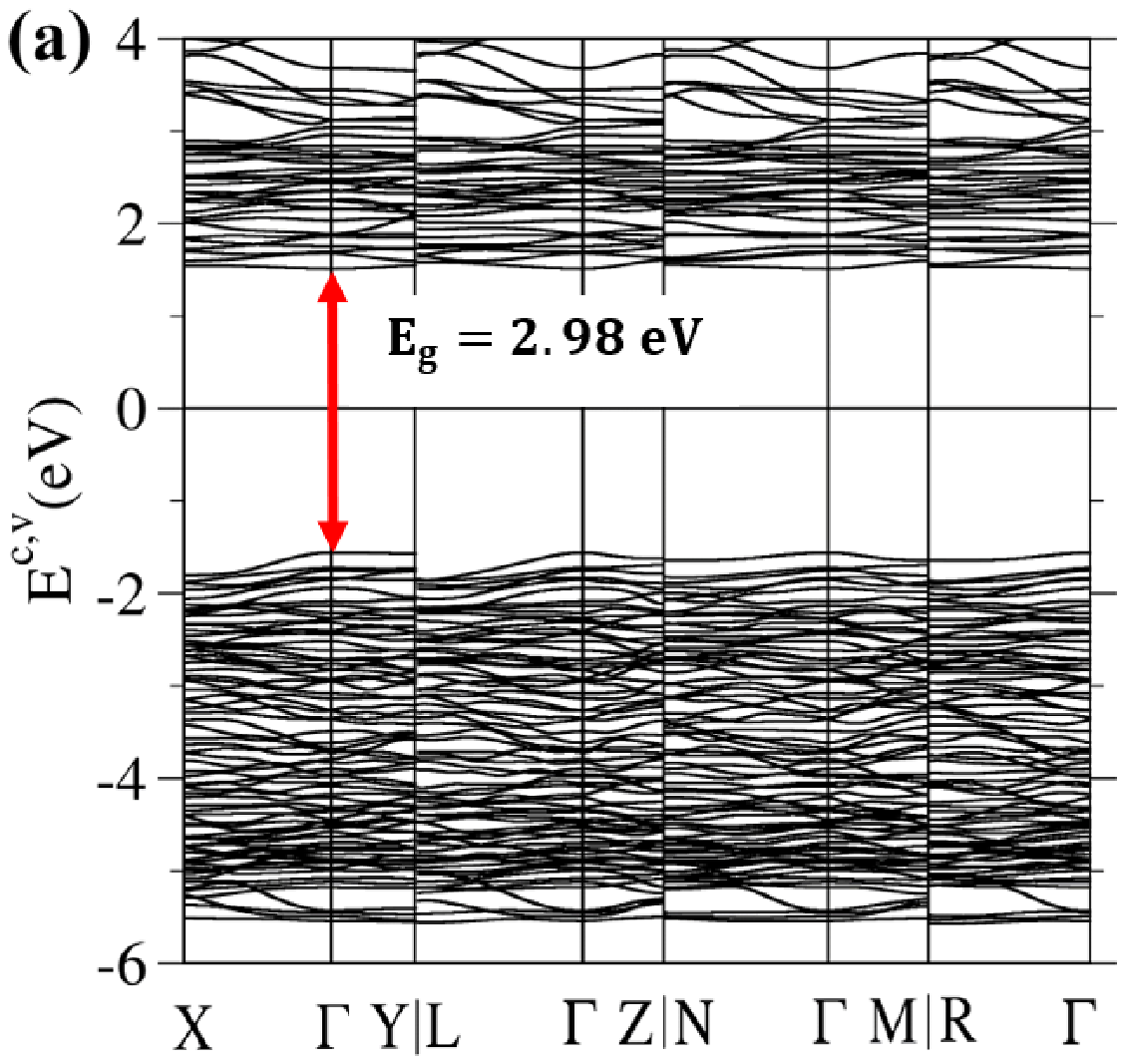}
            \includegraphics[scale=0.85]{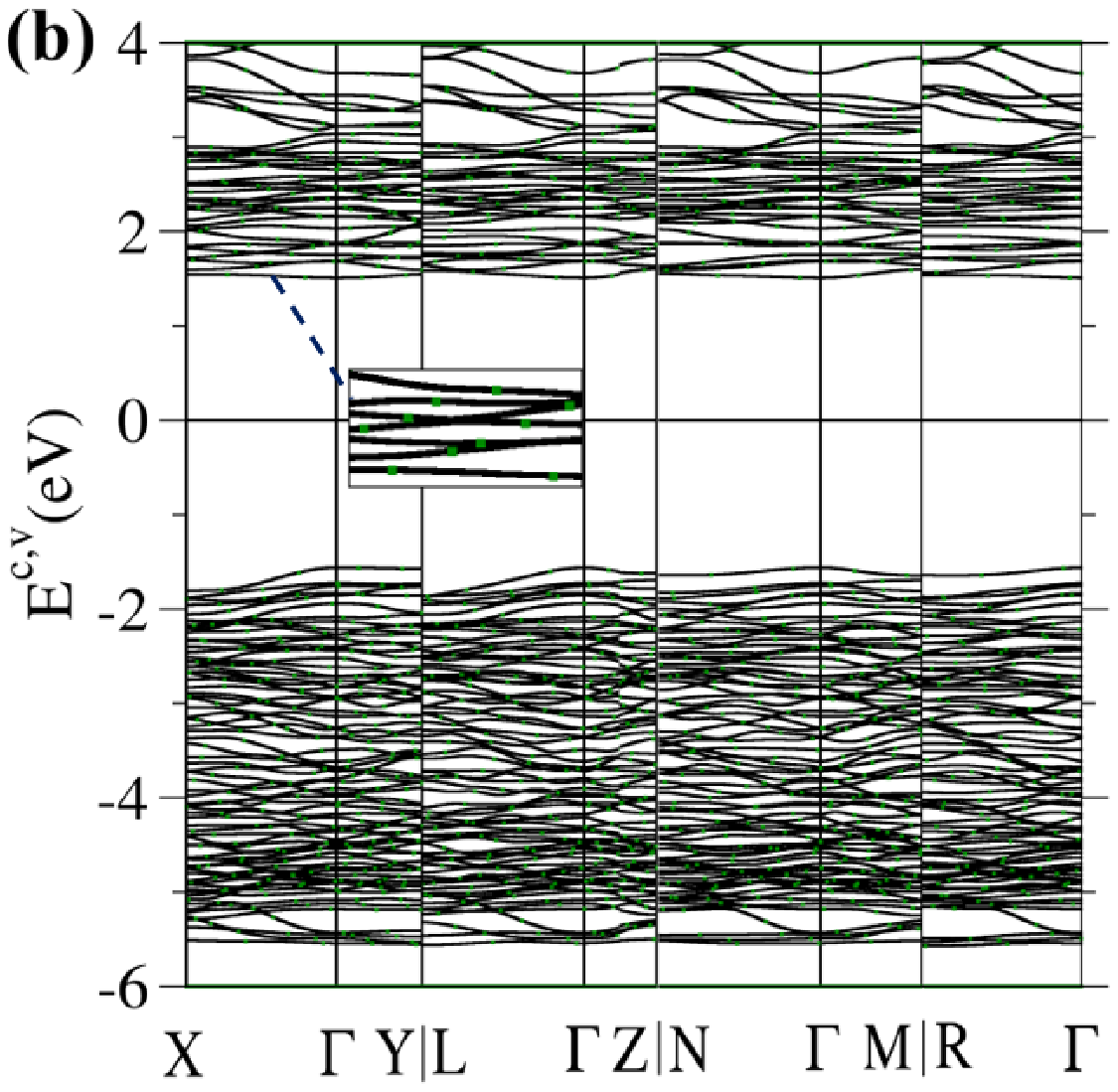}
\end{figure}
\vfill
\clearpage
\vfill
\begin{figure}[!ht]
            \centering
           \includegraphics[scale=0.85]{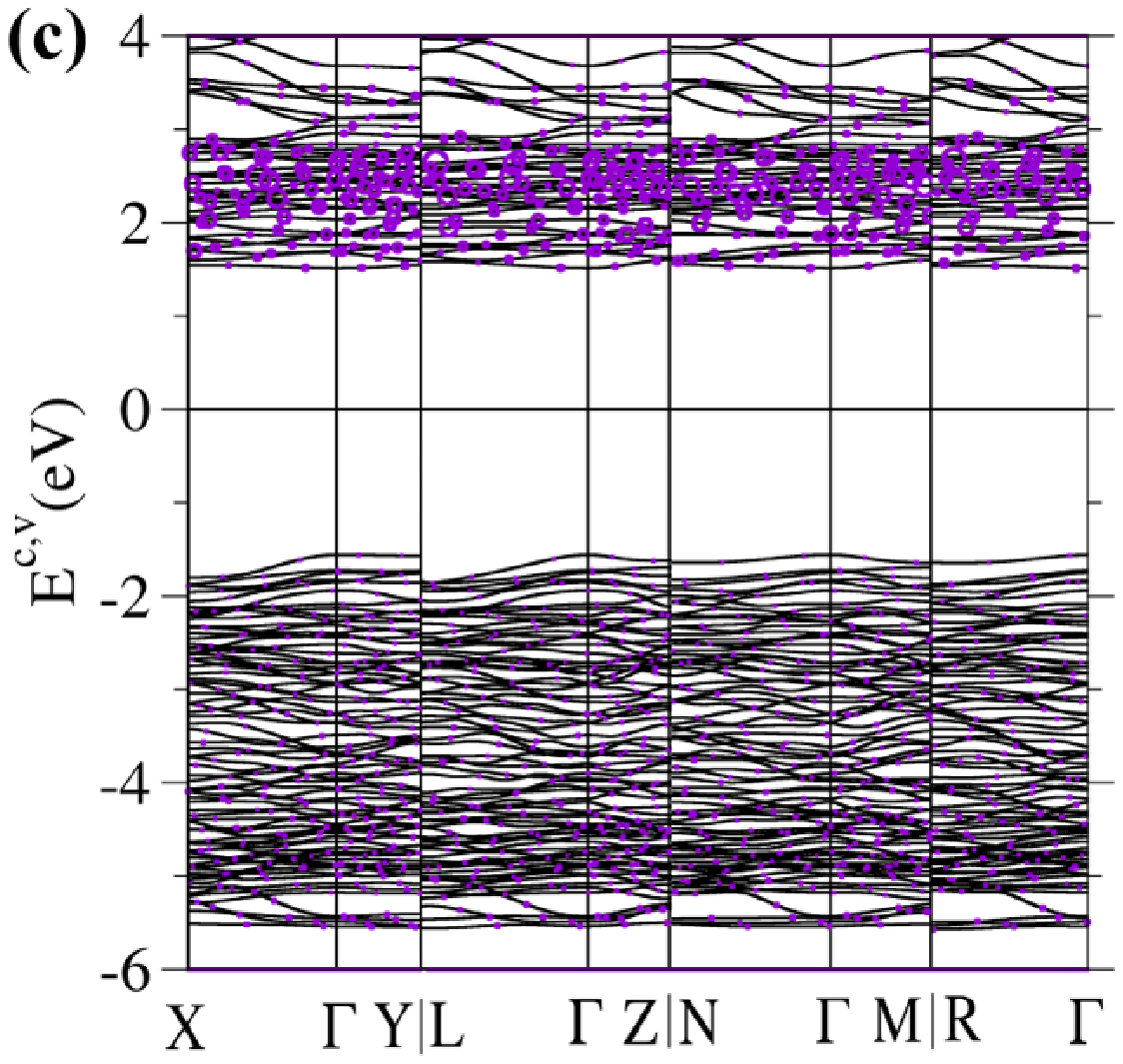}
           \centering
           \includegraphics[scale=0.85]{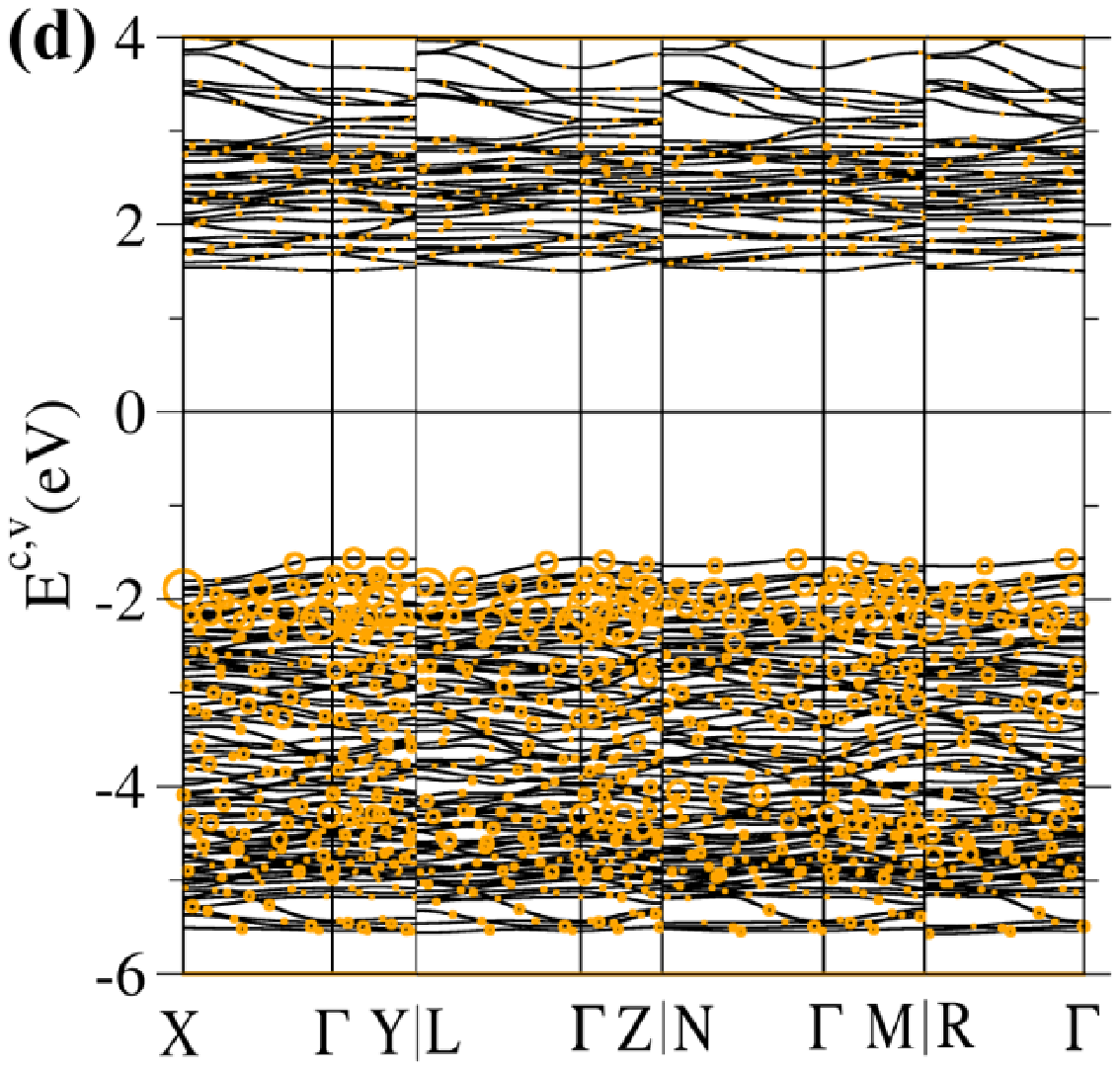}

\end{figure}

Figure 4.4: (a) Electronic energy spectrum for Li$_4$Ti$_5$O$_{12}$, along the high-symmetry points in the first Brillouin zone within the range ${-6.0}$ eV${\le\,E^v\le\,4.0}$ eV for the specific (b) lithium, (c) titanium and (d) oxygen dominances (green,violet and orange balls, respectively).
\vfill
\clearpage
\vfill
\begin{figure}[!ht]
\begin{center}
           \includegraphics[scale=0.7]{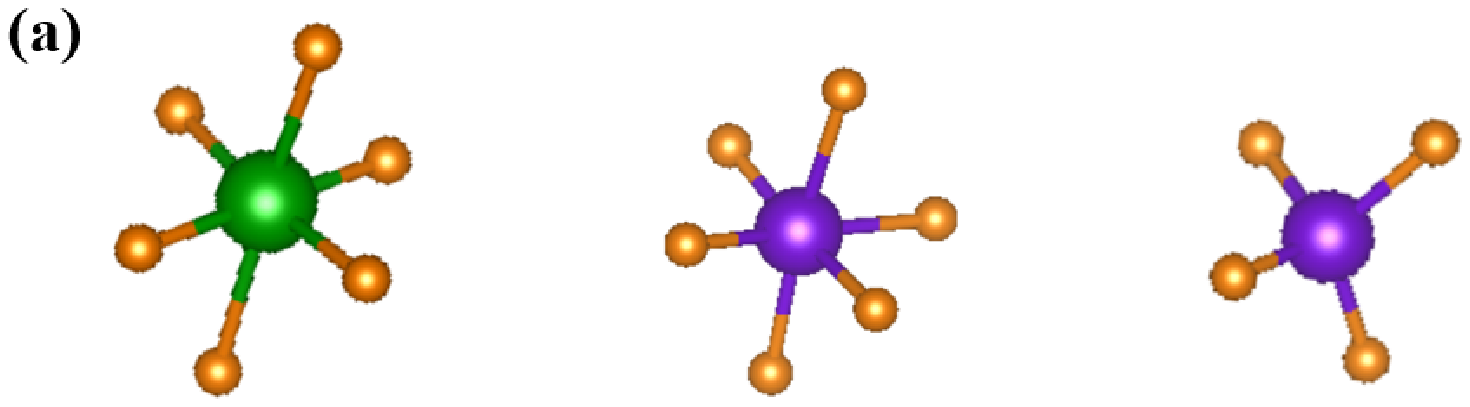}
           \includegraphics[scale=1.0]{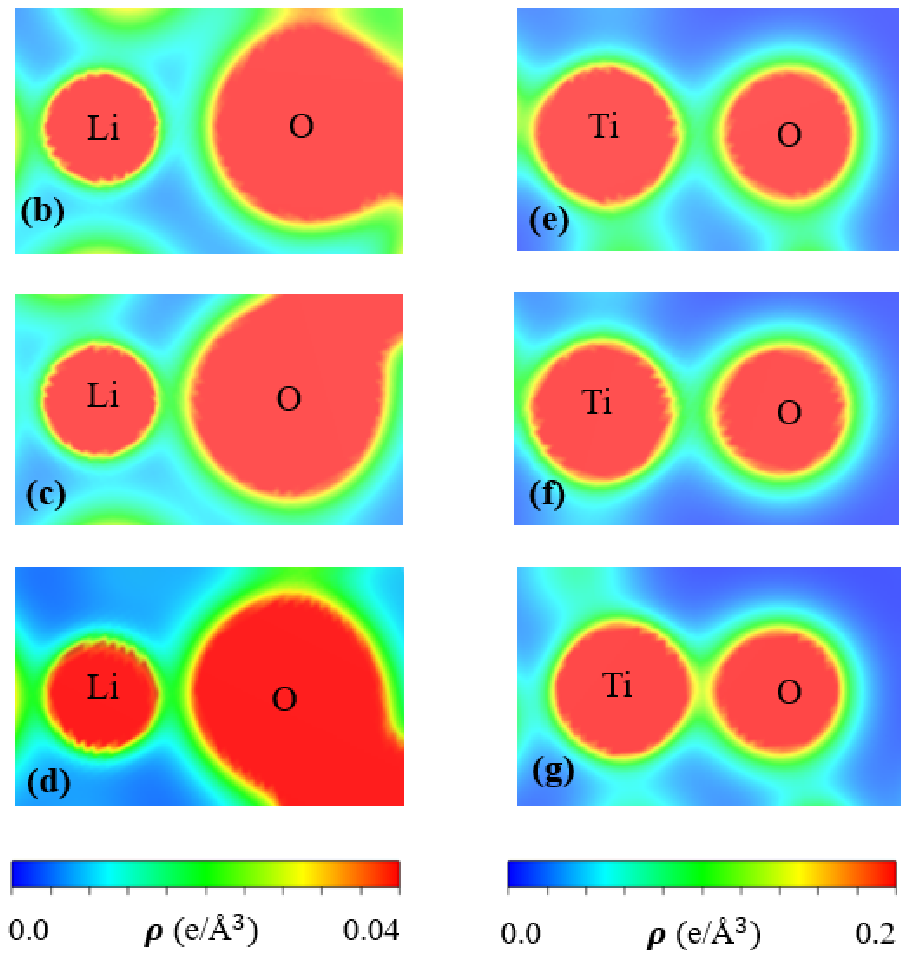}
           \end{center}
\end{figure}
\newpage
\begin{figure}[!ht]
\begin{center}
           \includegraphics[scale=0.7]{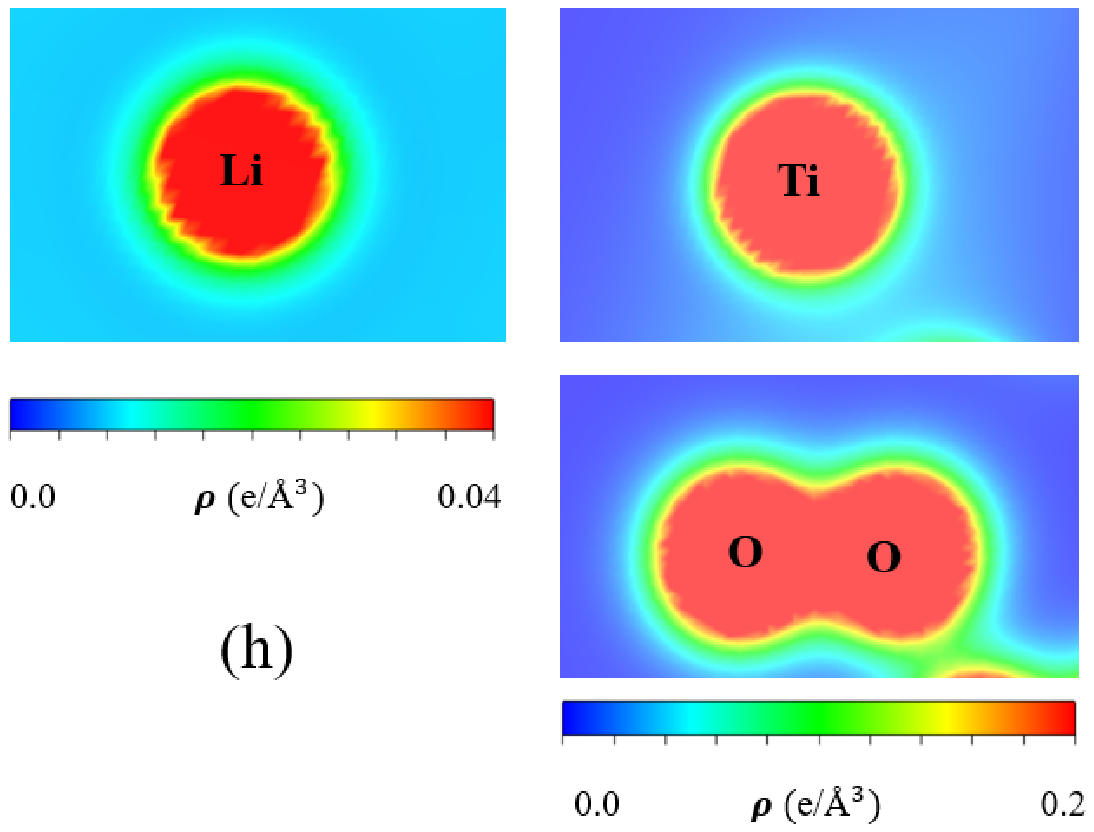}
           \end{center}
\end{figure}
Figure 4.5: (a) The different Li-O $\&$ Ti-O chemical bonds (LiO$_6$, TiO$_4$, TiO$_6$), in which the diversified spatial charge density distributions under the longest/middle/shortest for (b)/(c)/(d) the former and (e)/(f)/(g) the latter, respectively. Also shown in (h) are those of the isolated atoms.
\vfill
\clearpage
\vfill

\begin{figure}[!ht]
\begin{center}
           \includegraphics[scale=0.7]{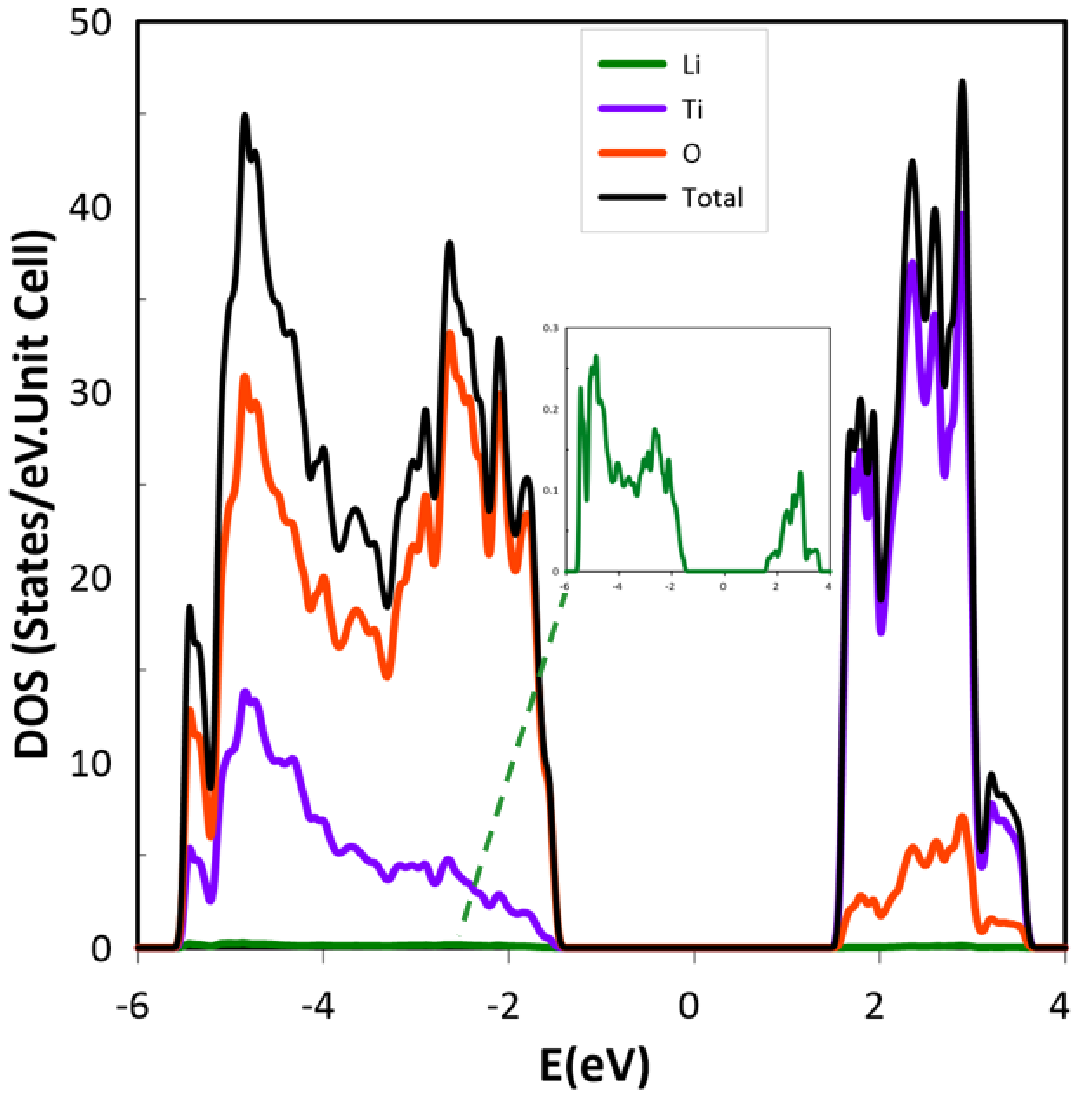}
       \end{center}
\end{figure}
\begin{figure}[!ht]
\begin{center}
 \includegraphics[scale=0.7]{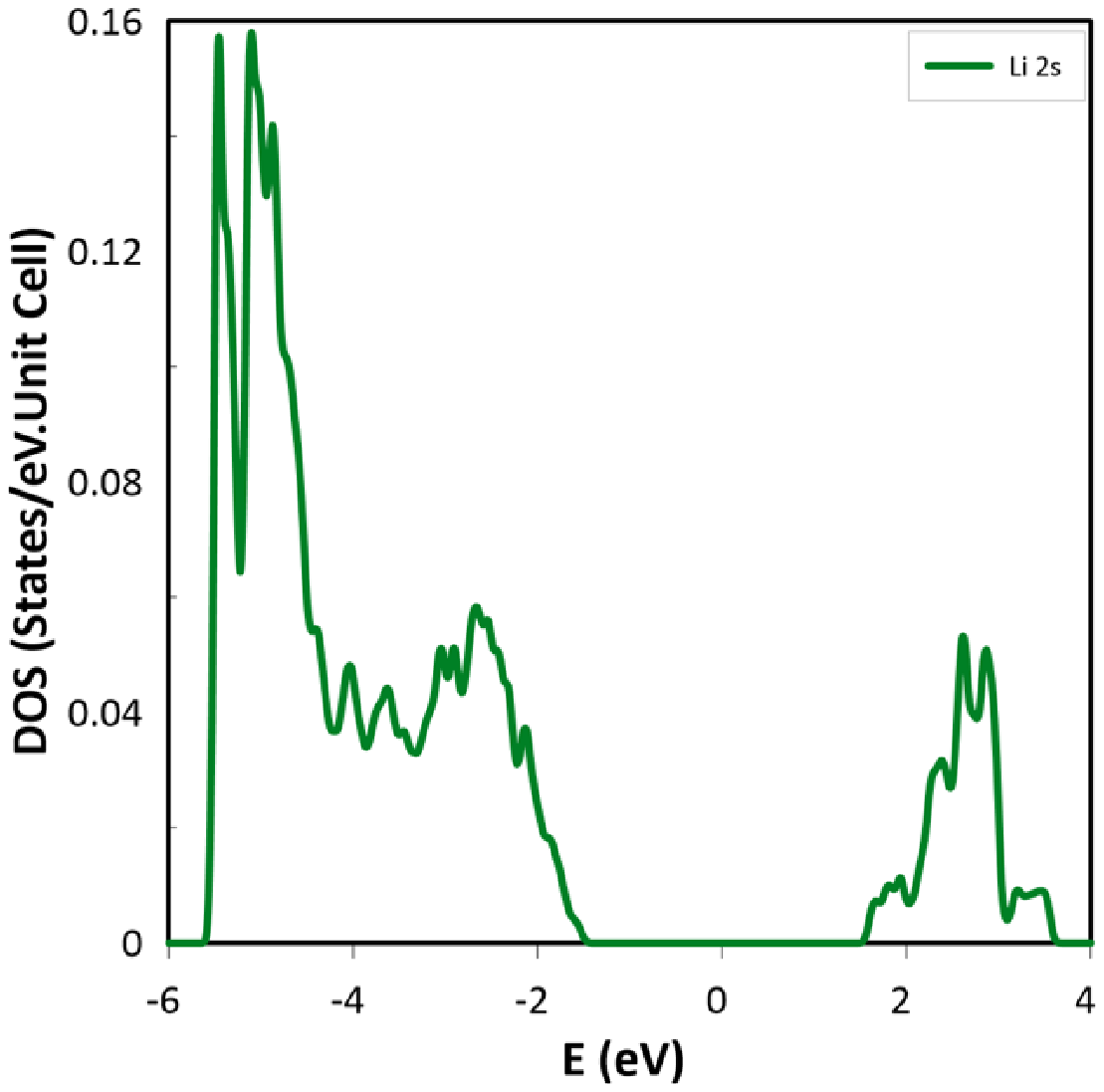}
 \end{center}
 \end{figure}
\vfill
\clearpage
    \begin{figure}[!ht]
    \begin{center}
          \includegraphics[scale=0.7]{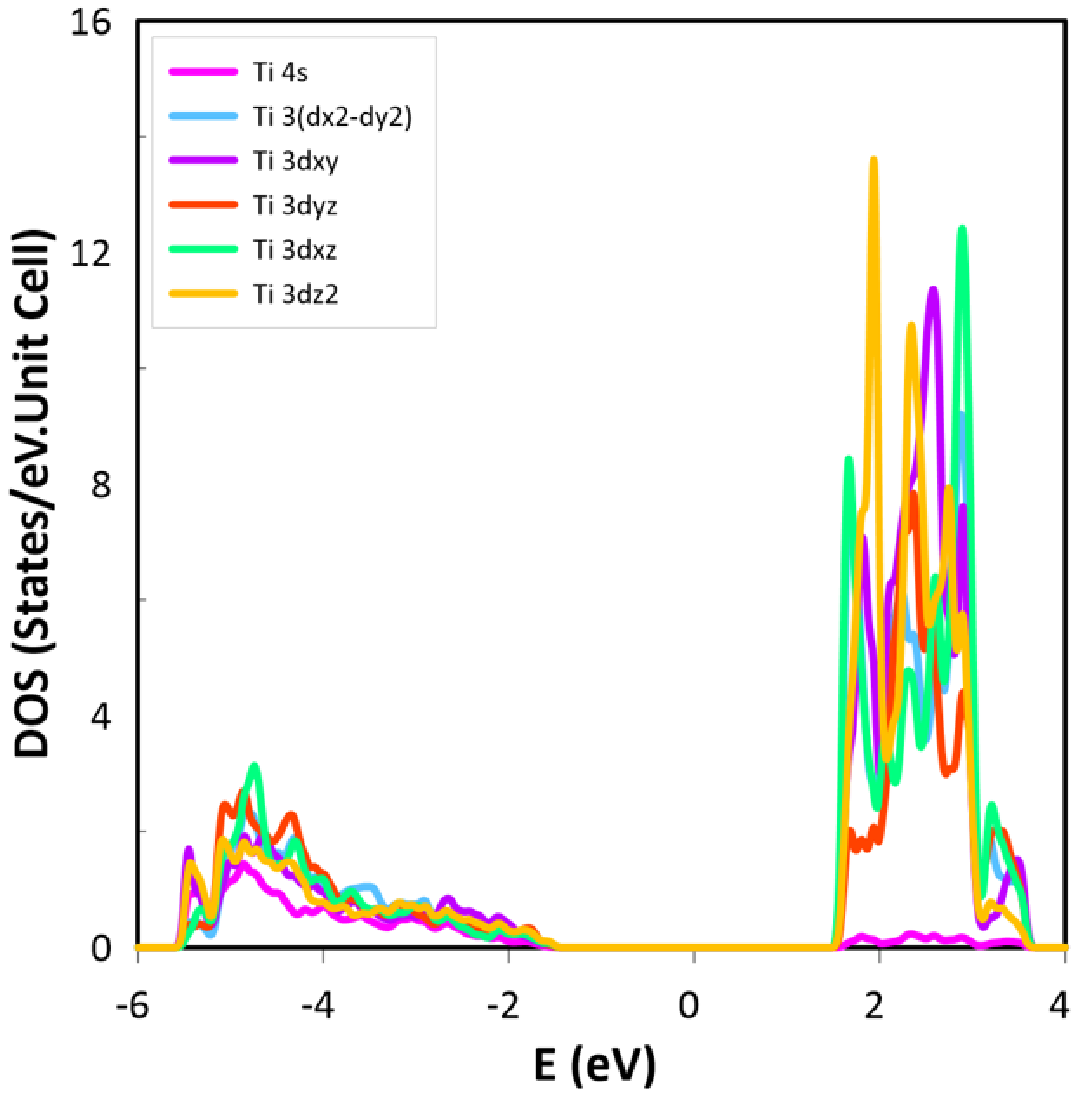}\\
           \includegraphics[scale=0.7]{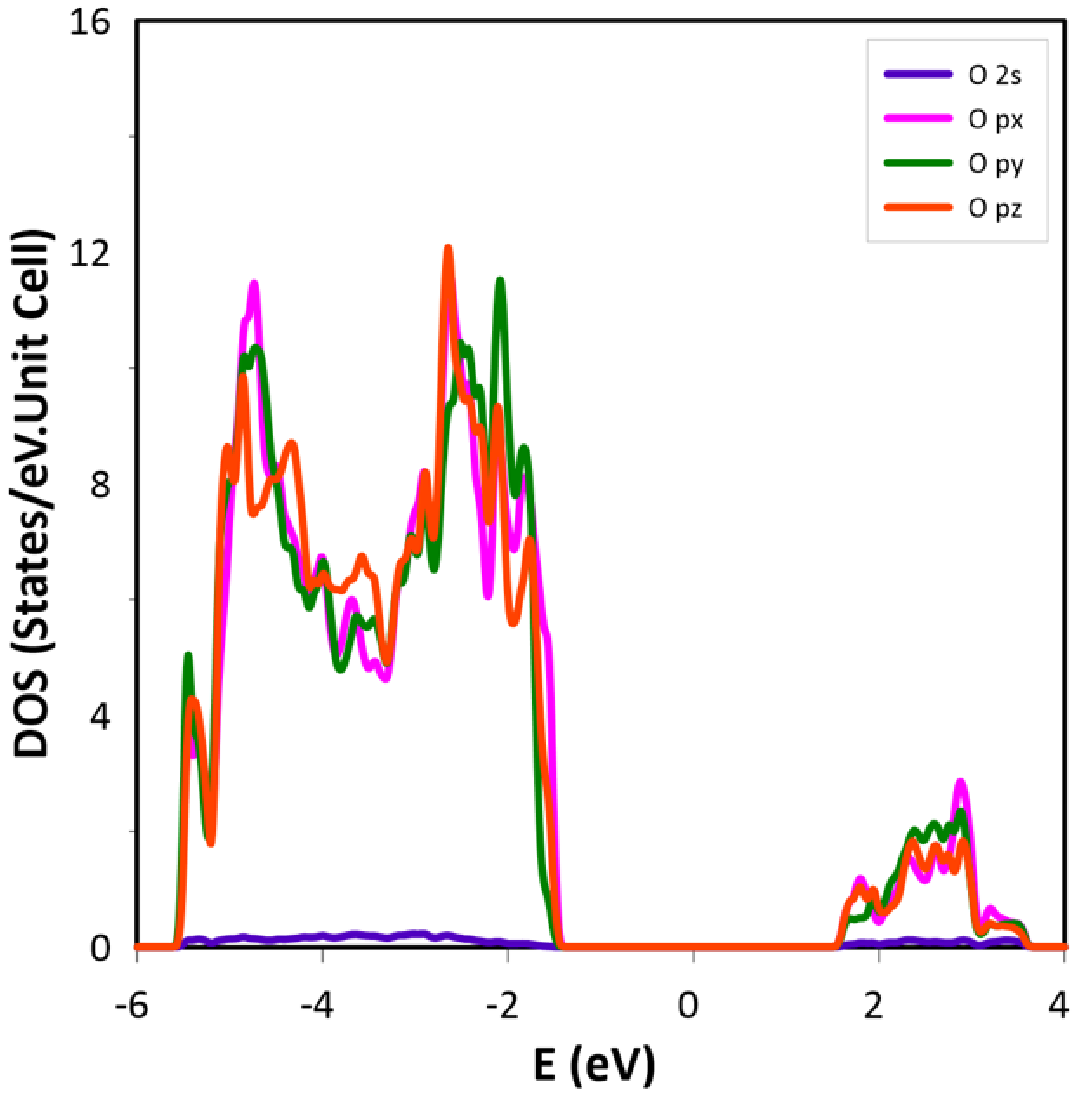}
           \end{center}
          \end{figure}
Figure 4.6: The atom- and orbital-decomposed density of states: those due to (a) Li, Ti, and O atoms (the green, violet and orange curves), (b) Li-2s orbital (the blue curve), (c) Ti-(4s, 3d${_{x^2-y^2}}$, 3d${_{xy}}$, 3d${_{yz}}$, 3d${_{zx}}$, 3d${_{z^2}}$) orbitals (the red, heavy red, purple, brown, light green; light blue curves), and (d) O-(2s, 2p$_x$, 2p$_y$, 2p$_z$) orbitals (the purple, pink, green and orange curves).

\end{document}